\documentclass[letterpaper, preprint, paper,11pt]{AAS}	

\usepackage{bm}
\usepackage{amsmath}
\usepackage{amssymb}
\usepackage{subfigure}
\usepackage[colorlinks=true, pdfstartview=FitV, linkcolor=black, citecolor= black, urlcolor= black]{hyperref}
\usepackage{overcite}
\usepackage{footnpag}			      	

\PaperNumber{XX-XXX}

\begin{document}

\title{Contraction Theory for Rendezvous and Proximity Operations}

\author{Suman Banerjee\thanks{Graduate Student, Department of Aerospace Engineering, University of Illinois Urbana-Champaign, 104 S. Wright St., Urbana, IL 61801.},
Tanayot Ngonsamrong\footnotemark[1],
Hrishav Das\footnotemark[1],
Veeresh Gudli\footnotemark[1],
Hiroyasu Tsukamoto\thanks{Assistant Professor, Department of Aerospace Engineering, University of Illinois Urbana-Champaign, 104 S. Wright St., Urbana, IL 61801.},
\ and Minduli Wijayatunga\footnotemark[2]
}

\maketitle{}

\begin{abstract}
This work investigates the disturbance tolerance on spacecraft and robust trajectory tracking for spacecraft rendezvous and proximity operations (RPO) during the final approach and docking phase. We combine trajectory optimization with contraction-based tracking guarantees to account for bounded disturbances while satisfying operational constraints. Sequential Convex Programming (SCP) is used to generate nominal docking trajectories under the Clohessy--Wiltshire relative-motion dynamics, subject to acceleration, fuel, terminal capture, and docking-cone constraints. A set of contraction metrics and feedback controllers is then used to construct certified tracking-error tubes around the nominal trajectories. Rather than approximating these tubes using a spherical bound, the proposed formulation preserves the directional geometry of the contraction-metric-induced error ellipsoid when tightening the docking-cone constraints. This allows the available tracking margin to depend on the direction of the constraint and provides a less conservative representation of the certified uncertainty. The resulting optimization determines the maximum disturbance magnitude that can be certified while maintaining the required approach constraints and prescribed nominal \(\Delta v\) budget, allowing the relationship between fuel usage and certified robustness to be evaluated. Directional constraint tightening is compared against spherical tube tightening and fixed-margin planning under identical disturbance conditions. Additionally, conformal prediction is used to construct a probabilistic tracking tube when disturbance data is available.

\end{abstract}

\section{Introduction}
Spacecraft RPO (Rendezvous and Proximity Operation) missions require a chaser to approach a target spacecraft placing several critical thresholds on approach distance, relative velocities and body orientation \cite{shahrezaei2026comprehensive}. Docking missions increase the severity of such constraints, requiring precise alignment in every parameter of both spacecraft as small guidance errors can prove to be very wasteful in terms of resources. In case of manned spaceflight, any error poses a direct threat to human safety and life-sustaining critical systems onboard the spacecraft. Therefore, during the final approach and docking phase, a planned trajectory must ensure that all safety parameters are strictly adhered to while keeping fuel consumption to a minimum. 
Recent literature shows the usefulness of Optimization theory in approaching the planning side of the problem. The CORTEX framework \cite{wijayatunga2026cortex,wijayatunga2025cortex} generates constrained trajectories for close-range rendezvous and uses convex optimization for tracking, with reference regeneration when the spacecraft deviates from its planned path and the final-approach strategy explicitly considers a docking corridor, keep-out region and operational limits. The planning of spacecraft rendezvous trajectories using successive convexification \cite{elango2026rendezvous}, a SCP-based real-time-capable solution method for nonconvex trajectory optimization addresses approach-cone constraints as well as solving nonconvex optimal control problems\cite{malyuta2022convex}. Spacecraft trajectory design and planning with avoidance constraints using mixed-interger linear programming methods have been developed to incorporate operational restrictions into an optimization framework\cite{richards2002spacecraft}. Together, these studies show the significance of constraint-aware optimization in RPO and motivate a closer examination of how tracking-error connects to constraints imposed on a nominal-docking trajectory.

Contraction theory provides us with an alternate analysis that asks a stronger question than conventional stability: Do arbitrary neighboring trajectories converge towards one another? Lohmiller et al. \cite{lohmiller1998contraction} talks about variational dynamics of non linear systems and analyzes where uniform negativity of the differential energy derivative yields exponential convergence between finite trajectories. This is analogous to Lyapunov Theory. Contraction theory builds on these results and subsequently the lyapunov theory to provide valuable guarantees on the bounds of deviations from nominal trajectories. Because of the guarantees contraction theory gives us, they have been used extensively in close tracking of trajectories and robust motion planning\cite{espindola2022new,singh2021safe,tsukamoto2021tutorial}. Zhao et al.\cite{zhao2022tube} develop robust control contraction metrics which guarantee invariant tubes to design trajectory controllers for nonlinear systems subject to external disturbances. CCM-based tubes have been further integrated into robust adaptive MPC to account for bounded uncertainty and disturbances under system constraints\cite{sasfi2023robust}.


This manuscript proposes a planning and control framework for the final approach and docking phase. Clohessy Wiltshire (CW) equations\cite{clohessy1960terminal} of relative motion are considered here. Sequential Convex Programming (SCP) will generate a nominal relative-motion trajectory that balances total \(\Delta v\)  and terminal docking accuracy while satisfying thrust and docking-cone constraints. A feedback controller based on a valid CCM for the docker spacecraft will then track that nominal trajectory. Under the conditions required by the contraction analysis, bounded disturbances would lead to exponential convergence toward a bounded tracking-error tube. This bound could also inform how much clearance the nominal trajectory needs from the docking-cone boundary. A tracking bound alone does not establish constraint satisfaction: the planned trajectory must leave enough margin for the resulting error set throughout the approach.

\section{Model Considered and Preliminaries}

\subsection{A. Dynamics}

In this work, the target spacecraft is assumed to follow a circular orbit about
the Earth, and the chaser operates in close proximity to the target, such that
the relative distance is much smaller than the target's orbital radius. Under
these assumptions, the relative motion of the chaser with respect to the target
is described by the Clohessy--Wiltshire (CW) equations, expressed in the
target's local-vertical local-horizontal (LVLH) frame.

The state vector is defined as
$\mathbf{x} = [\mathbf{r}^{\top}\;\mathbf{v}^{\top}]^{\top}$, where
$\mathbf{r}=[r_1,r_2,r_3]^{\top}$ and $\mathbf{v}=[v_1,v_2,v_3]^{\top}$ denote
the relative position and velocity of the chaser, with components along the
radial, along-track, and cross-track directions, respectively. The CW dynamics
form a linear time-invariant system
\begin{equation}
\dot{\mathbf{x}} = \mathbf{A}\mathbf{x} + \mathbf{B}\mathbf{u},
\label{eq:cw_dynamics}
\end{equation}
with
\begin{equation}
\mathbf{A} =
\begin{bmatrix}
0 & 0 & 0 & 1 & 0 & 0\\
0 & 0 & 0 & 0 & 1 & 0\\
0 & 0 & 0 & 0 & 0 & 1\\
3n^2 & 0 & 0 & 0 & 2n & 0\\
0 & 0 & 0 & -2n & 0 & 0\\
0 & 0 & -n^2 & 0 & 0 & 0
\end{bmatrix},
\qquad
\mathbf{B} =
\begin{bmatrix}
\mathbf{0}_{3\times3}\\
\mathbf{I}_{3}
\end{bmatrix},
\end{equation}
where $n=\sqrt{\mu_E/a_c^3}$ is the mean motion of the target, $\mu_E$ is Earth's gravitational
parameter, and $a_c$ is the chief orbit radius and $\mathbf{u}\in\mathbb{R}^3$ is
the control acceleration applied to the chaser.

\subsection{B. Sequential Convex Programming (SCP)}
The objective of this work is to generate docking trajectories from a range of
initial states that minimize control effort under the CW dynamics. For a fixed
time of flight $t_f$, the trajectory optimization problem is formulated as \cite{chari2024spacecraftrendezvousguidancefactorizationfree}
\begin{equation}
\begin{aligned}
\min_{\mathbf{x}(\cdot),\,\mathbf{u}(\cdot)} \quad
  & \int_{0}^{t_f} \|\mathbf{u}(t)\|_2^2 \, dt \\
\text{subject to} \quad
  & \dot{\mathbf{x}}(t) = \mathbf{A}\mathbf{x}(t) + \mathbf{B}\mathbf{u}(t), \\
  & \mathbf{x}(0) = \mathbf{x}_0, \qquad
    \mathbf{x}(t_f) = \mathbf{x}_{\text{dock}}, \\
  & \|\mathbf{u}(t)\|_2 \leq u_{\max}, \\
  & \|\mathbf{r}(t) - \mathbf{r}_c\|_2 \geq \rho_c(t),
    \qquad \forall t \in [0, t_f], \\
  & \hat{\mathbf{d}}^{\top}(\mathbf{r}(t)-\mathbf{r}_p) \geq
    \cos\theta_a\,\|\mathbf{r}(t)-\mathbf{r}_p\|_2,
    \qquad \forall t \in [t_f - t_a,\, t_f], \\
  & \hat{\mathbf{e}}(t)^{\top}\mathbf{u}(t) \leq
    \cos\theta_p\,\|\mathbf{u}(t)\|_2,
    \qquad \forall t \in [t_f - t_a,\, t_f],
\end{aligned}
\label{eq:docking_ocp}
\end{equation}
where $u_{\max}$ is the maximum available control acceleration, and the
keep-out-zone (KOZ) constraint requires the chaser to remain at least a distance
$\rho_c(t)$ from the target center $\mathbf{r}_c$. The KOZ radius is held at its
nominal value $\rho_c$ until the final approach phase of duration $t_a$, after
which it decreases monotonically to a terminal value
$\rho_c(t_f) \leq \|\mathbf{r}_p - \mathbf{r}_c\|_2$, where $\mathbf{r}_p$ is the
docking port position, i.e., the position component of $\mathbf{x}_{\text{dock}}$.
During this phase, the chaser is additionally constrained to an approach
corridor, which requires the vector from the docking port to the chaser to lie
within an angle $\theta_a$ of the outward docking direction $\hat{\mathbf{d}}$,
so that the KOZ is relaxed only along the docking axis. The last constraint
addresses plume impingement. Since the thruster exhaust is directed along
$-\mathbf{u}$, braking during the final approach would otherwise direct the
plume toward the target, causing contamination, heating, and disturbance forces
on the target. The constraint therefore requires the thrust direction to make an
angle of at least $\theta_p$ with the line of sight from the target to the
chaser,
$\hat{\mathbf{e}}(t) = (\mathbf{r}(t)-\mathbf{r}_c)/\|\mathbf{r}(t)-\mathbf{r}_c\|_2$,
so that the exhaust does not enter a cone of half-angle $\theta_p$ about the
target direction.

Since the CW dynamics are linear and the control and corridor constraints are
second-order cones, the only nonconvex elements of
Eq.~\eqref{eq:docking_ocp} are the KOZ and plume impingement constraints. The
problem is discretized over $N$ intervals of length $\Delta t = t_f/N$ using a
zero-order hold, which for linear time-invariant dynamics yields the exact
discrete-time model
$\mathbf{x}_{k+1} = \mathbf{A}_d \mathbf{x}_k + \mathbf{B}_d \mathbf{u}_k$, with
$\mathbf{A}_d = e^{\mathbf{A}\Delta t}$ and
$\mathbf{B}_d = \int_0^{\Delta t} e^{\mathbf{A}\tau}\,d\tau\,\mathbf{B}$.
Under the same hold, the cost becomes
$\sum_{k=0}^{N-1}\|\mathbf{u}_k\|_2^2$. The final approach phase
corresponds to the node set $\mathcal{K}_a = \{N-N_a,\dots,N\}$ with
$N_a = t_a/\Delta t$, and the KOZ radius at node $k$ is denoted
$\rho_{c,k} = \rho_c(k\Delta t)$.

Let $(\bar{\mathbf{x}}^{(j)}, \bar{\mathbf{u}}^{(j)}) =
\{\bar{\mathbf{x}}_k^{(j)}, \bar{\mathbf{u}}_k^{(j)}\}$ denote the reference
trajectory at SCP iteration $j$, given by the solution of the previous
iteration or, for $j=0$, by an initial guess. The KOZ constraint is linearized
about the reference positions $\bar{\mathbf{r}}_k^{(j)}$, giving the supporting
half-spaces
\begin{equation}
\hat{\mathbf{n}}_k^{(j)\top}(\mathbf{r}_k - \mathbf{r}_c) \geq \rho_{c,k},
\qquad
\hat{\mathbf{n}}_k^{(j)} = \frac{\bar{\mathbf{r}}_k^{(j)} - \mathbf{r}_c}
                          {\|\bar{\mathbf{r}}_k^{(j)} - \mathbf{r}_c\|_2}.
\label{eq:koz_linearized}
\end{equation}
Because $\|\mathbf{r}-\mathbf{r}_c\|_2$ is convex, each half-space is an inner
approximation of the KOZ-feasible set, so any solution of the subproblem
satisfies the original KOZ constraint at the nodes.

The plume impingement constraint is convexified in a similar manner. The line
of sight is evaluated at the reference,
$\hat{\mathbf{e}}_k^{(j)} = (\bar{\mathbf{r}}_k^{(j)}-\mathbf{r}_c)/
\|\bar{\mathbf{r}}_k^{(j)}-\mathbf{r}_c\|_2$, and $\|\mathbf{u}_k\|_2$ is
replaced by its linearization about the reference control
$\bar{\mathbf{u}}_k^{(j)}$, giving
\begin{equation}
\hat{\mathbf{e}}_k^{(j)\top}\mathbf{u}_k \leq
\cos\theta_p\,\frac{\bar{\mathbf{u}}_k^{(j)\top}\mathbf{u}_k}
                    {\|\bar{\mathbf{u}}_k^{(j)}\|_2}.
\label{eq:plume_linearized}
\end{equation}
Since $\|\mathbf{u}\|_2 \geq \bar{\mathbf{u}}^{\top}\mathbf{u}/\|\bar{\mathbf{u}}\|_2$,
any $\mathbf{u}_k$ satisfying Eq.~\eqref{eq:plume_linearized} also satisfies
the original cone constraint for the fixed line of sight
$\hat{\mathbf{e}}_k^{(j)}$, which coincides with the true line of sight upon
convergence. At nodes where $\bar{\mathbf{u}}_k^{(j)} = \mathbf{0}$, the
right-hand side is replaced by zero, which is conservative. The approach
corridor is a second-order cone constraint and is imposed directly without
linearization. The convex subproblem solved at iteration $j$ is then
\begin{equation}
\begin{aligned}
\min_{\{\mathbf{x}_k\},\,\{\mathbf{u}_k\}} \quad
  & \sum_{k=0}^{N-1} \|\mathbf{u}_k\|_2^2 \\
\text{subject to} \quad
  & \mathbf{x}_{k+1} = \mathbf{A}_d\mathbf{x}_k + \mathbf{B}_d\mathbf{u}_k,
    && k = 0,\dots,N-1, \\
  & \mathbf{x}_0 = \mathbf{x}_0, \qquad
    \mathbf{x}_N = \mathbf{x}_{\text{dock}}, \\
  & \|\mathbf{u}_k\|_2 \leq u_{\max},
    && k = 0,\dots,N-1, \\
  & \hat{\mathbf{n}}_k^{(j)\top}(\mathbf{r}_k - \mathbf{r}_c) \geq \rho_{c,k},
    && k = 0,\dots,N, \\
  & \hat{\mathbf{d}}^{\top}(\mathbf{r}_k-\mathbf{r}_p) \geq
    \cos\theta_a\,\|\mathbf{r}_k-\mathbf{r}_p\|_2,
    && k \in \mathcal{K}_a, \\
  & \hat{\mathbf{e}}_k^{(j)\top}\mathbf{u}_k \leq
    \cos\theta_p\,\bar{\mathbf{u}}_k^{(j)\top}\mathbf{u}_k/
    \|\bar{\mathbf{u}}_k^{(j)}\|_2,
    && k \in \mathcal{K}_a \setminus \{N\}, \\
  & \|\mathbf{x}_k - \bar{\mathbf{x}}_k^{(j)}\|_2 \leq \delta_x,
    && k = 0,\dots,N, \\
  & \|\mathbf{u}_k - \bar{\mathbf{u}}_k^{(j)}\|_2 \leq \delta_u,
    && k = 0,\dots,N-1,
\end{aligned}
\label{eq:scp_subproblem}
\end{equation}

The initial state $\mathbf{x}_0$ is prescribed and is not
an optimization variable. The last two constraints form a trust region that keeps each iterate
close to the reference, where the linearizations are accurate. Problem
\eqref{eq:scp_subproblem} is a second-order cone program (SOCP). Its solution
becomes the reference $(\bar{\mathbf{x}}^{(j+1)}, \bar{\mathbf{u}}^{(j+1)})$
for the next iteration, and the process is repeated until
$\max_k\|\mathbf{x}_k - \bar{\mathbf{x}}_k^{(j)}\|_2 < \epsilon_{\text{tol}}$,
yielding a locally optimal docking trajectory
$(\mathbf{x}_d, \mathbf{u}_d)$.

\subsection{C. Contraction Theory}
The nominal dynamics of the rendezvous and proximity operations (RPO) mission
are given by the CW model in Eq.~\eqref{eq:cw_dynamics}. However, the actual
system may be subject to unknown disturbances and modeling uncertainties, such
as higher-order gravitational terms, differential drag, solar radiation
pressure, and actuator errors. Tube-based model predictive control (MPC)
achieves robustness to such effects by maintaining the state within a robust
positively invariant set around the nominal trajectory, typically by solving an
optimization problem online. Contraction theory instead certifies incremental
exponential stability through a differential condition on a metric
$\mathbf{M}(\mathbf{x},t)$. The resulting bound holds for any desired
trajectory, provides an explicit convergence rate and tube radius in closed
form, and requires no online optimization beyond evaluating a feedback law.
The same condition also extends directly to time-varying and nonlinear
dynamics, where invariant-set computations become considerably more difficult.

The desired trajectory $(\mathbf{x}_d(t),\mathbf{u}_d(t))$ is obtained from the
SCP solution of Eq.~\eqref{eq:docking_ocp}, with $\mathbf{u}_d(t)$ applied as a
zero-order hold. The chaser tracks the desired trajectory using the feedback law
\begin{equation}
\mathbf{u}(t) = \mathbf{u}_d(t) + \mathbf{K}\big(\mathbf{x}(t)-\mathbf{x}_d(t)\big),
\label{eq:tracking_law}
\end{equation}
which yields the closed-loop nominal dynamics
\begin{equation}
f(\mathbf{x},t) = \mathbf{A}\mathbf{x}
  + \mathbf{B}\big(\mathbf{u}_d(t) + \mathbf{K}(\mathbf{x}-\mathbf{x}_d(t))\big),
\label{eq:closed_loop}
\end{equation}
with Jacobian $\partial f/\partial\mathbf{x} = \mathbf{A}+\mathbf{B}\mathbf{K}$.
Since the zero-order-hold discretization in Sec.~B is exact,
$\mathbf{x}_d(t)$ is a solution of $\dot{\mathbf{x}} = f(\mathbf{x},t)$.

Consider the perturbed closed-loop dynamics
\begin{equation}
    \dot{\mathbf{x}} = f(\mathbf{x},t) + \mathbf{d}(\mathbf{x},t),
    \label{eq:perturbed_dynamics}
\end{equation}
where $\mathbf{d}(\mathbf{x},t)$ represents an unknown but bounded disturbance
satisfying
\begin{equation}
    \sup_{\mathbf{x},t} \left\| \mathbf{d}(\mathbf{x},t) \right\| \leq \bar{d} < \infty,
    \label{eq:disturbance_bound}
\end{equation}
where $\bar{d}$ denotes the maximum disturbance magnitude.
Let $\mathbf{x}(t)$ denote the actual perturbed trajectory. The desired and
perturbed trajectories are connected by a smooth one-parameter family of
trajectories $\mathbf{q}(\mu,t)$, where $\mu \in [0,1]$ is a path parameter
satisfying
\begin{equation}
\mathbf{q}(0,t) = \mathbf{x}_d(t),
\qquad
\mathbf{q}(1,t) = \mathbf{x}(t).
\label{eq:path_bc}
\end{equation}
The virtual system is defined along the path $\mathbf{q}(\mu,t)$ as
\begin{equation}
\dot{\mathbf{q}}
=
f(\mathbf{q},t)
+
\mu \mathbf{d}(\mathbf{x},t),
\qquad \mu \in [0,1],
\label{eq:virtual_system}
\end{equation}
so that $\mathbf{q}(0,t)$ follows the nominal closed-loop dynamics and
$\mathbf{q}(1,t)$ follows the perturbed dynamics. Differentiating
Eq.~\eqref{eq:virtual_system} with respect to $\mu$ gives the differential
dynamics
\begin{equation}
\frac{d}{dt}
\left(
\frac{\partial \mathbf{q}}{\partial \mu}
\right)
=
\frac{\partial f}{\partial \mathbf{q}}
\frac{\partial \mathbf{q}}{\partial \mu}
+
\mathbf{d}(\mathbf{x},t),
\label{eq:differential_dynamics}
\end{equation}
where $\partial f/\partial\mathbf{q}$ is the Jacobian of the closed-loop
dynamics evaluated along $\mathbf{q}(\mu,t)$.

Suppose there exists a uniformly positive-definite contraction metric
$\mathbf{M}(\mathbf{x},t)$ satisfying
\begin{equation}
m\mathbf{I}
\preceq
\mathbf{M}(\mathbf{x},t)
\preceq
\bar{m}\mathbf{I},
\qquad
\forall \mathbf{x},t,
\label{eq:metric_bounds}
\end{equation}
where $0 < m \leq \bar{m}$, and the Riemannian distance induced by
$\mathbf{M}$ is defined as
\begin{equation}
d_M(t)
=
\inf_{\gamma \in \Gamma_t}
\int_0^1
\sqrt{
\left(\frac{\partial \gamma}{\partial \mu}\right)^{\top}
\mathbf{M}(\gamma(\mu),t)
\left(\frac{\partial \gamma}{\partial \mu}\right)
}
\, d\mu,
\label{eq:riemannian_distance}
\end{equation}
where $\Gamma_t$ denotes the set of piecewise smooth paths satisfying
$\gamma(0)=\mathbf{x}_d(t)$ and $\gamma(1)=\mathbf{x}(t)$. The same integral
evaluated along the particular path $\mathbf{q}(\mu,t)$ defines its Riemannian
length, which upper-bounds $d_M(t)$, with equality when $\mathbf{q}(\mu,t)$ is
a length-minimizing geodesic.

If the continuous-time contraction condition
\begin{equation}
\dot{\mathbf{M}}
+
\mathbf{M}\frac{\partial f}{\partial \mathbf{x}}
+
\left(\frac{\partial f}{\partial \mathbf{x}}\right)^{\top}
\mathbf{M}
\preceq
-2\alpha\mathbf{M},
\qquad
\forall \mathbf{x},t,\quad \alpha>0,
\label{eq:contraction_condition}
\end{equation}
holds, then the Riemannian distance between the perturbed and desired
trajectories satisfies
\begin{equation}
d_M(t)
\leq
d_M(0)e^{-\alpha t}
+
\frac{\bar{d}\sqrt{\bar{m}}}{\alpha}
\left(1-e^{-\alpha t}\right),
\label{eq:contraction_error_bound}
\end{equation}
and consequently the state-space tracking error is bounded by
\begin{equation}
    \left\lVert
        \mathbf{x}(t) - \mathbf{x}_d(t)
    \right\rVert
    \leq
    \frac{d_M(0)}{\sqrt{m}}e^{-\alpha t}
    +
    \frac{\bar{d}}{\alpha}
    \sqrt{\frac{\bar{m}}{m}}
    \left(1-e^{-\alpha t}\right).
    \label{eq:state_error_bound}
\end{equation}
The tracking error therefore converges exponentially to a tube of radius
$(\bar{d}/\alpha)\sqrt{\bar{m}/m}$ around the desired trajectory.

For the linear closed-loop system in Eq.~\eqref{eq:closed_loop}, a constant
metric $\mathbf{M}$ suffices, so that $\dot{\mathbf{M}}=\mathbf{0}$ and
Eq.~\eqref{eq:contraction_condition} reduces to
\begin{equation}
(\mathbf{A}+\mathbf{B}\mathbf{K})^{\top}\mathbf{M}
+ \mathbf{M}(\mathbf{A}+\mathbf{B}\mathbf{K})
\preceq -2\alpha\mathbf{M}.
\label{eq:contraction_lmi}
\end{equation}
The construction of a metric $\mathbf{M}$ and a gain $\mathbf{K}$ satisfying
this condition is described next.

\subsubsection{D. Feedback Gain from the Contraction Metric}

In Eq.~\eqref{eq:tracking_law}, the feedback gain $\mathbf{K}$ is generic. In
this work, the feedback is instead constructed directly from the contraction
metric, so that designing $\mathbf{M}$ simultaneously designs the controller.
Defining the tracking error $\mathbf{e}(t) = \mathbf{x}(t)-\mathbf{x}_d(t)$,
the tracking law is chosen as
\begin{equation}
\mathbf{u}(t) = \mathbf{u}_d(t)
  - \mathbf{R}^{-1}\mathbf{B}^{\top}\mathbf{M}\,\mathbf{e}(t),
\qquad \text{i.e.,} \qquad
\mathbf{K} = -\mathbf{R}^{-1}\mathbf{B}^{\top}\mathbf{M},
\label{eq:metric_feedback}
\end{equation}
where $\mathbf{R} \succ 0$ is a control weighting matrix. This form of
metric-based feedback is standard in the contraction
literature~\cite{manchester2017ccm,tsukamoto2021tutorial}. To interpret it,
let $\mathbf{u}_{fb} = \mathbf{u}-\mathbf{u}_d$ denote the feedback term, so
that the nominal tracking error evolves as
$\dot{\mathbf{e}} = \mathbf{A}\mathbf{e} + \mathbf{B}\mathbf{u}_{fb}$. The
feedback in Eq.~\eqref{eq:metric_feedback} is then the unique minimizer of
\begin{equation}
\frac{d}{dt}\big(\mathbf{e}^{\top}\mathbf{M}\mathbf{e}\big)
+ \mathbf{u}_{fb}^{\top}\mathbf{R}\,\mathbf{u}_{fb}
\label{eq:feedback_interpretation}
\end{equation}
over $\mathbf{u}_{fb}$, and therefore balances the rate of decrease of the
Riemannian energy $\mathbf{e}^{\top}\mathbf{M}\mathbf{e}$ against the weighted
control effort.

Substituting Eq.~\eqref{eq:metric_feedback} into
Eq.~\eqref{eq:contraction_lmi}, pre- and post-multiplying by
$\mathbf{W} = \mathbf{M}^{-1}$, and noting that
$\mathbf{B}\mathbf{K}\mathbf{W} = -\mathbf{B}\mathbf{R}^{-1}\mathbf{B}^{\top}$,
the contraction condition becomes
\begin{equation}
\mathbf{A}\mathbf{W} + \mathbf{W}\mathbf{A}^{\top}
- 2\,\mathbf{B}\mathbf{R}^{-1}\mathbf{B}^{\top} + 2\alpha\mathbf{W}
\preceq \mathbf{0}.
\label{eq:ccm_condition}
\end{equation}
The control weighting is parameterized as $\mathbf{R} = \mathbf{R}_0/s$, where
$\mathbf{R}_0 \succ 0$ specifies the relative weighting between control axes
and $s > 0$ is a scalar that sets the overall feedback strength.
Eq.~\eqref{eq:ccm_condition} then reads
\begin{equation}
\mathbf{A}\mathbf{W} + \mathbf{W}\mathbf{A}^{\top}
- 2s\,\mathbf{B}\mathbf{R}_0^{-1}\mathbf{B}^{\top} + 2\alpha\mathbf{W}
\preceq \mathbf{0},
\label{eq:ccm_lmi}
\end{equation}
which is a linear matrix inequality in $(\mathbf{W},s)$. Since the pair
$(\mathbf{A},\mathbf{B})$ of the CW dynamics is controllable,
Eq.~\eqref{eq:ccm_lmi} is feasible for any prescribed contraction rate
$\alpha > 0$.

Because Eq.~\eqref{eq:ccm_lmi} is invariant under a joint scaling of
$(\mathbf{W},s)$, which leaves $\mathbf{K}$ unchanged, the scale of the metric
can be fixed without loss of generality by imposing
$\mathbf{I} \preceq \mathbf{W} \preceq \chi\mathbf{I}$. This implies
$m = 1/\chi$ and $\bar{m} = 1$ in Eq.~\eqref{eq:metric_bounds}, so that
$\sqrt{\bar{m}/m} = \sqrt{\chi}$. The metric is then computed from the
semidefinite program
\begin{equation}
\begin{aligned}
\min_{\mathbf{W},\,s,\,\chi} \quad & \chi + \lambda s \\
\text{subject to} \quad
  & \mathbf{A}\mathbf{W} + \mathbf{W}\mathbf{A}^{\top}
    - 2s\,\mathbf{B}\mathbf{R}_0^{-1}\mathbf{B}^{\top} + 2\alpha\mathbf{W}
    \preceq \mathbf{0}, \\
  & \mathbf{I} \preceq \mathbf{W} \preceq \chi\mathbf{I}, \qquad s \geq 0,
\end{aligned}
\label{eq:ccm_sdp}
\end{equation}
where $\lambda \geq 0$ trades off the tube size against the feedback
magnitude. The solution yields the contraction metric
$\mathbf{M} = \mathbf{W}^{-1}$, the control weighting
$\mathbf{R} = \mathbf{R}_0/s$, the feedback gain
$\mathbf{K} = -\mathbf{R}^{-1}\mathbf{B}^{\top}\mathbf{M}$, and the tube radius
$(\bar{d}/\alpha)\sqrt{\chi}$ in Eq.~\eqref{eq:state_error_bound}. Moreover,
since $\|\mathbf{M}\| \leq 1$, the feedback effort satisfies
$\|\mathbf{K}\mathbf{e}\| \leq s\,\|\mathbf{R}_0^{-1}\|\,\|\mathbf{e}\|$,
which can be checked against the control margin below $u_{\max}$.

\subsubsection{E. Conformal Prediction}
Conformal Prediction is a statistical distribution-free framework used to generate finite-sample coverage guarantees for arbitrary uncertainty predictors based on the rank statistics of non-conformity scores. Given a regression predictor $\hat{f}$ trained independently of a calibration dataset
$\mathcal{D}_{\mathrm{cal}}=\{(x_i,y_i)\}_{i=1}^{N_{\rm cal}}$,
define the nonconformity scores $s_i$ as
\begin{equation}
    s_i = \left\|y_i-\hat{f}(x_i)\right\|,
    \qquad i=1,\ldots,N_{\rm cal},
    \label{eq:cp_nonconformity_score}
\end{equation}
where $N_{\rm cal}$ denotes the number of calibration samples. For a prescribed failure probability $\delta \in (0,1)$, let the conformal quantile be denoted as $q_{1-\delta}$ which is the $\left\lceil (N_{\rm cal}+1)(1-\delta)\right\rceil$-th smallest nonconformity score in $\{s_i\}_{i\in\{1,2,..,N_{\rm cal}\}}$. For a new sample $(X_{i_{new}},Y_{i_{new}})$, if the nonconformity scores $\{s_i\}_{i \in \{1,2,..,N_{\rm cal}\}\cup\{i_{new}\}}$ are exchangeable, then the following statistical guarantee holds:
\begin{equation}
    \mathbb{P}\left(\|\hat{f}(X_{i_{new}})-Y_{i_{new}}\|\leq q_{1-\delta}\right)\geq 1-\delta,
    \label{eq:cp_basic_coverage}
\end{equation}
where $\mathbb{P}$ is the probability measure induced by the joint distribution of the data $\{(X_i,Y_i)\}_{i=1}^{N_{\rm cal}}$ and the new samples $(X_{i_{\mathrm{new}}},Y_{i_{\mathrm{new}}})$.

\begin{figure}
    \centering
    \includegraphics[width=1\linewidth]{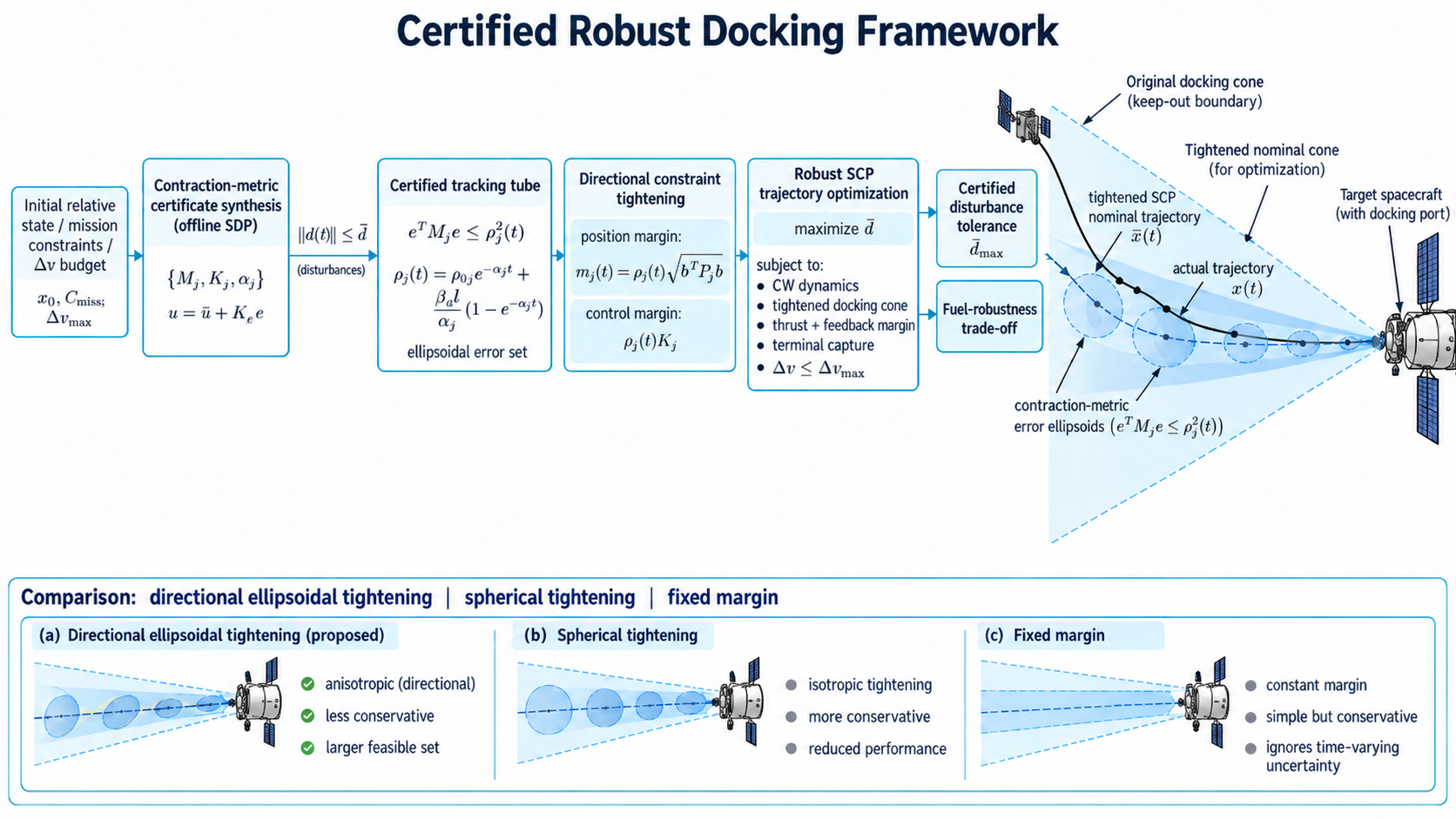}
    \caption{Contraction based Tracking for Docking}
    \label{fig:placeholder}
\end{figure}

\section{Methodology}

Certifying how much disturbance a docking maneuver can tolerate requires more than a well-designed nominal trajectory or a stable tracking controller on its own; the two must be designed with each other in mind. The methodology in this work therefore integrates SCP-based trajectory planning with contraction-metric tracking, so that disturbance tolerance can be certified directly against the docking constraints. A nominal docking trajectory is first computed under the CW dynamics and is then tracked by a contraction-based controller, which, for bounded disturbances, yields a time-varying tracking-error tube around the nominal path. Instead of bounding this tube with a sphere, the docking-cone constraints are tightened using its directional geometry, so that the margin reserved along each direction reflects how far the error can actually grow in that direction. The disturbance limit is then increased until the tightened constraints can no longer be satisfied, and the largest feasible value is taken as the certified disturbance tolerance. Finally, when disturbance data are available, conformal prediction is used to construct a statistical disturbance envelope, producing a probabilistic tracking tube.

\subsection{Nominal Trajectory Design}

The first step is to generate a nominal state-control trajectory
$(\mathbf{x}_d(t),\mathbf{u}_d(t))$ that brings the chaser to the
docking port while respecting the operational limits of the mission. This trajectory is obtained via SCP under the CW equations
\begin{equation}
\dot{\mathbf{x}}_d = \mathbf{A}\mathbf{x}_d
+ \mathbf{B}\mathbf{u}_d,
\qquad \mathbf{x}_d(0)=\mathbf{x}_0,
\label{eq:nominal_cw_dynamics}
\end{equation}
subject to thrust, keep-out-zone, terminal capture, and other constraints. Unlike a fixed target point, the docking port moves with the target spacecraft, so its position and orientation must be tracked in the LVLH frame. With the port located at body-frame position $\ell_p$ and oriented along the body-frame axis $\hat{\mathbf{d}}_B$, it moves according to $r_p(t)=\mathbf{R}_t(t)\boldsymbol{\ell}_p$ and $\hat{\mathbf{d}}(t)=R_t(t)\hat{\mathbf{d}}_B$, where $R_t(t)$ is the rotation from the target body frame to the LVLH frame. The nominal offset of the chaser from the port is then $\mathbf{s}_d=\mathbf{r}_d-r_p(t)$. To ensure that the chaser approaches along the port axis, a docking cone of half-angle $\theta_a$ is enforced through a family of supporting half-spaces with normals
\begin{equation}
\mathbf{b}(\boldsymbol{\nu},t)
=\sin\theta_a\,\hat{\mathbf{d}}(t)
-\cos\theta_a\,\boldsymbol{\nu},
\qquad
\boldsymbol{\nu}^\top\hat{\mathbf{d}}(t)=0,
\;\|\boldsymbol{\nu}\|=1.
\label{eq:docking_cone_normal}
\end{equation}
The nominal docking-cone constraint is therefore
\begin{equation}
\mathbf{b}(\boldsymbol{\nu},t)^\top
[\mathbf{r}_d(t)-\mathbf{r}_p(t)]\ge0,
\qquad
\forall\boldsymbol{\nu}:
\boldsymbol{\nu}^\top\hat{\mathbf{d}}(t)=0,
\;\|\boldsymbol{\nu}\|=1,
\label{eq:nominal_docking_cone}
\end{equation}
which keeps the nominal path inside the cone throughout the approach.

\subsection{Contraction-Based Controller Design}

A nominal trajectory by itself offers no protection against disturbances, since any deviation from the plan is left uncorrected over the rest of the maneuver. The second step is therefore to design a tracking controller whose behavior can be certified. For a chosen contraction rate $\alpha$ and weight $\lambda$, we solve the SDP in Eq.~\eqref{eq:ccm_sdp} to obtain a constant metric $\mathbf{M}\succ0$ together with the feedback gain
$\mathbf{K}=-\mathbf{R}^{-1}\mathbf{B}^\top\mathbf{M}$. Defining the tracking error as $\mathbf{e}=\mathbf{x}-\mathbf{x}_d$,
the closed-loop control is
\begin{equation}
\mathbf{u} = \mathbf{u}_d + \mathbf{K}\mathbf{e}.
\label{eq:method_tracking_control}
\end{equation}
A useful consequence of the linearity of the CW model is that the same pair $(\mathbf{M},\mathbf{K})$ applies to any nominal trajectory, so the controller needs to be designed only once. Repeating the synthesis for multiple values of $\alpha$ (and the corresponding weight $\lambda$) provides us with a set of candidate certificates $(\mathbf{M}_j,\mathbf{K}_j,\alpha_j)$, each of which satisfies
\begin{equation}
(\mathbf{A}+\mathbf{B}\mathbf{K}_j)^\top\mathbf{M}_j
+\mathbf{M}_j(\mathbf{A}+\mathbf{B}\mathbf{K}_j)
\preceq -2\alpha_j\mathbf{M}_j.
\label{eq:candidate_contraction_certificate}
\end{equation}

\subsection{Tracking-Tube Construction and Constraint Tightening}

The certificates above describe how quickly tracking errors decay, but to connect them with the docking constraints we must also describe how large those errors can become. Uncertainties are modeled as a matched disturbance $\mathbf{d}(t)$ entering through the control channel,
\begin{equation}
\dot{\mathbf{x}} = \mathbf{A}\mathbf{x}
+ \mathbf{B}\mathbf{u} + \mathbf{B}\mathbf{d},
\qquad \|\mathbf{d}(t)\|\le \bar d.
\label{eq:matched_disturbance_dynamics}
\end{equation}
For each certificate $(M_j,K_j,\alpha_j)$, let $\rho_{0j}=\sup_{\mathbf{e}_0\in\mathcal{E}_0}
\sqrt{\mathbf{e}_0^\top\mathbf{M}_j\mathbf{e}_0}$ denote the largest initial error measured in the metric over the initial error set $\mathcal{E}_0$, and let $\beta_j=\|\mathbf{M}_j^{1/2}\mathbf{B}\|_2$ describe how strongly the disturbance enters in the same metric. Contraction then guarantees that
\begin{equation}
\begin{aligned}
\mathbf{e}(t)^\top\mathbf{M}_j\mathbf{e}(t)
&\le \rho_j^2(t),\\
\rho_j(t) &= \rho_{0j}e^{-\alpha_j t}
+\frac{\beta_j\bar d}{\alpha_j}\bigl(1-e^{-\alpha_j t}\bigr).
\end{aligned}
\label{eq:tracking_tube_bound}
\end{equation}
so the tracking error remains inside an ellipsoid whose size transitions exponentially from $\rho_{0j}$ to the steady-state value $\beta_j\bar d/\alpha_j$.

Since the docking cone constrains only the position of the chaser, this ellipsoid is projected onto position using $\mathbf{P}_j=\mathbf{C}_r\mathbf{M}_j^{-1}\mathbf{C}_r^\top$,
where $\mathbf{C}_r=[\mathbf{I}_3\ \mathbf{0}_{3\times3}]$. The largest positional error along a constraint normal $\mathbf{b}=\mathbf{b}(\boldsymbol{\nu},t)$ is then $\rho_j(t)\sqrt{\mathbf{b}^\top\mathbf{P}_j\mathbf{b}}$, which depends on the direction of $b$. We thus tighten the docking-cone condition to
\begin{equation}
\mathbf{b}(\boldsymbol{\nu},t)^\top\mathbf{s}_d(t)
\ge \rho_j(t)\sqrt{
\mathbf{b}(\boldsymbol{\nu},t)^\top
\mathbf{P}_j\mathbf{b}(\boldsymbol{\nu},t)}
\quad
\forall\boldsymbol{\nu}:
\boldsymbol{\nu}^\top\hat{\mathbf{d}}(t)=0,
\;\|\boldsymbol{\nu}\|=1,
\label{eq:cone_tightening}
\end{equation}

which guarantees that the entire error ellipsoid, and not only the nominal position, stays inside the cone. A spherical bound, by contrast, would replace the directional term with $\rho_j\sqrt{\lambda_{\max}(\mathbf{P}_j)}$ and reserve the worst-case margin in every direction. The same reasoning applies to the thrust limit. The feedback acceleration is bounded by $\|\mathbf{K}_j\mathbf{e}\|\le\rho_j(t)\kappa_j$
with $\kappa_j=\|\mathbf{K}_j\mathbf{M}_j^{-1/2}\|_2$, so we enforce $\|\mathbf{u}_d(t)\|_2+\rho_j(t)\kappa_j\le u_{\max}$ to leave enough control authority for the feedback to act.

With these tightened constraints in place, the disturbance bound $\bar d$ is treated as a decision variable rather than a fixed parameter. An SCP trajectory optimization is solved to maximize $\bar d$ subject to the CW dynamics, thrust and fuel limits, capture and keep-out constraints, and the tightened cone constraint above. 

For prescribed final time $t_f$ and nominal $\Delta v$ budget
$\Delta v_{\max}$, solve
\begin{equation}
\begin{aligned}
 \max_{j,\bar d,\mathbf{x}_d,\mathbf{u}_d}\quad&\bar d\\
 \text{subject to}\quad&
 \dot{\mathbf{x}}_d=\mathbf{A}\mathbf{x}_d
 +\mathbf{B}\mathbf{u}_d,\quad
 \mathbf{x}_d(0)=\mathbf{x}_0,\\
 &\bar d\geq0,\quad
 \|\mathbf{u}_d(t)\|_2+\rho_j(t)\kappa_j\leq u_{\max},\\
 &\text{Eq.~\eqref{eq:cone_tightening}},\qquad
 \int_0^{t_f}\|\mathbf{u}_d(t)\|_2\,dt\leq\Delta v_{\max},\\
 &\text{terminal capture and other operational constraints}.
\end{aligned}
\label{eq:disturbance_tolerance}
\end{equation}
Here $u_{\max}$ is the acceleration limit. Terminal constraints
reserve tracking margins for specified position and relative-velocity
tolerances at a finite capture region. The selected certificate is
held fixed during execution. The optimum measures disturbance
tolerance certified by the candidate set, not the spacecraft's
absolute robustness limit. Varying $\Delta v_{\max}$ gives a
certified fuel--robustness trade-off.


The result is the largest $\bar d$ for which all constraints remain feasible, which we refer to as the certified disturbance tolerance. Solving this problem for each $(\mathbf{M}_j,\mathbf{K}_j)$ and for various $\Delta v_{\max}$ allows directional tightening to be compared against spherical tightening and fixed-margin approaches under identical conditions.

\subsection{Conformal Prediction for Disturbance Calibration}

The certified tolerance above assumes that a deterministic bound $\bar d$ is known in advance, which is rarely the case in practice, since perturbations such as differential drag and solar radiation pressure vary with the mission conditions. When such a bound is unavailable, conformal prediction can instead calibrate it from data. Given $N_{\rm cal}$ sample disturbances $\mathbf{d}_i(t)$ recorded under conditions $z_i$, we choose an acceleration scale $\sigma_d(z,t)>0$ and compute the nonconformity scores
\begin{equation}
S_i=\sup_{t\in[0,t_f]}
\frac{\|\mathbf{d}_i(t)\|_2}{\sigma_d(z_i,t)},
\qquad i=1,\ldots,N_{\rm cal},
\label{eq:cp_disturbance_score}
\end{equation}
each of which measures the worst normalized disturbance over one sample. Let $q_{1-\delta}$ be the $(1-\delta)$-quantile of the $S_i$, where $\delta \in (0,1)$. Define the calibrated disturbance envelope as $\bar d_{\rm CP}(z,t)=q_{1-\delta}\,\sigma_d(z,t)$. Then, by exchangeability,
\begin{equation}
\mathbb{P}\bigl(\|\mathbf{d}_{\rm new}(t)\|\le\bar d_{\rm CP}(z_{\rm new},t)
\ \forall t\in[0,t_f]\bigr)\ge1-\delta.
\label{eq:cp_disturbance_coverage}
\end{equation}
For the selected certificate, we omit the index $j$ and write
$\mathbf{M}=\mathbf{M}_j$, $\alpha=\alpha_j$, $\beta=\beta_j$,
$\rho_0=\rho_{0j}$, and $\mathbf{P}=\mathbf{P}_j$. Propagating $\bar d_{\rm CP}(t)$ through the contraction bound in place of the constant $\bar d$ yields a probabilistic tube
\begin{equation}
\rho_{\rm CP}(t)=\rho_0e^{-\alpha t}
+ \beta\int_0^t e^{-\alpha(t-s)}\,\bar d_{\rm CP}(s)\,ds,
\label{eq:cp_tracking_tube}
\end{equation}
so that
\begin{equation}
\mathbb{P}\bigl(
\mathbf{e}(t)^\top\mathbf{M}\mathbf{e}(t)
\le\rho_{\rm CP}^2(t)
\ \forall t\in[0,t_f]\bigr)\ge1-\delta.
\label{eq:cp_tube_coverage}
\end{equation}
The docking cone is then tightened by $\rho_{\rm CP}(t)$ exactly as before (e.g.\ $\mathbf{b}^\top\mathbf{s}_d
\ge\rho_{\rm CP}\sqrt{\mathbf{b}^\top\mathbf{P}\mathbf{b}}$). A spherical alternative uses $\epsilon_r^{\rm CP}(t)=\rho_{\rm CP}(t)
\|\mathbf{C}_r\mathbf{M}^{-1/2}\|_2$ with
\begin{equation}
h(\mathbf{r}_d,t)=\hat{\mathbf{d}}(t)^\top[\mathbf{r}_d-\mathbf{r}_p(t)]
-\cos\theta_a\,\|\mathbf{r}_d-r_p(t)\|
\ge(1+\cos\theta_a)\,\epsilon_r^{\rm CP}(t).
\label{eq:cp_spherical_tightening}
\end{equation}
It should be noted that this coverage guarantee holds only when the calibration trajectories and the new mission trajectories are exchangeable.

\section{Expected Results}

To evaluate the proposed framework, a dataset of nominal docking trajectories
will be generated by solving the SCP problem in
Eq.~\eqref{eq:scp_subproblem} from a large set of initial states
$\mathbf{x}_0$ sampled over the approach region. The contraction
metric $\mathbf{M}$ and the feedback gain
$\mathbf{K} = -\mathbf{R}^{-1}\mathbf{B}^{\top}\mathbf{M}$ will be computed
once from the semidefinite program in Eq.~\eqref{eq:ccm_sdp}. Since the
closed-loop Jacobian $\mathbf{A}+\mathbf{B}\mathbf{K}$ does not depend on the
nominal trajectory, the same metric and gain apply to every trajectory in the
dataset, and no per-trajectory controller design is required.

Each nominal trajectory will then be tracked in closed-loop simulation using
the feedback law in Eq.~\eqref{eq:metric_feedback}, in the presence of
disturbances that are unknown to the controller, such as higher-order
gravitational perturbations, differential drag, solar radiation pressure, and
actuator errors. Monte Carlo simulations with randomized disturbance
realizations will be performed, and the resulting tracking error
$\|\mathbf{e}(t)\|$ will be compared with the theoretical bound in
Eq.~\eqref{eq:state_error_bound}. Note that the controller itself does not
require knowledge of the disturbance; the disturbance bound $\bar{d}$ enters
only the value of the guaranteed error bound.

It is expected that, for all nominal trajectories and all disturbance
realizations satisfying Eq.~\eqref{eq:disturbance_bound}, the tracking error
remains bounded and converges exponentially to the tube of radius
$(\bar{d}/\alpha)\sqrt{\chi}$ around the desired trajectory. Since the chaser
starts on the nominal trajectory, $d_M(0)=0$, and the error is expected to
remain within this tube for the entire maneuver. This behavior is in contrast
to open-loop execution of the nominal control $\mathbf{u}_d$, for which the
along-track drift of the CW dynamics causes the tracking error to grow without
bound. The results are also expected to quantify the trade-off governed by the
contraction rate $\alpha$ and the weight $\lambda$ in
Eq.~\eqref{eq:ccm_sdp}: faster contraction and smaller $\chi$ yield tighter
tubes at the cost of larger feedback effort, which will be compared with the
available control margin below $u_{\max}$ and with the KOZ clearance of the
nominal trajectories.

Finally, because the tube radius is derived from the worst-case disturbance
bound $\bar{d}$, it is expected to be conservative, with the observed tracking
errors lying well inside the tube in most realizations. To address this, the
tracking errors from the Monte Carlo simulations will be used as calibration
data for conformal prediction, yielding a statistically calibrated tube radius
that contains the tracking error of a new realization with probability at least
$1-\delta$. This calibrated radius is expected to be substantially smaller
than the worst-case bound while retaining a distribution-free probabilistic
guarantee, enabling less conservative constraint tightening for safe docking.

\section{Conclusion}
This paper proposes a framework for robust spacecraft docking under the CW dynamics that combines sequential convex programming, contraction theory, and conformal prediction. SCP generates locally optimal docking trajectories that respect a shrinking keep-out zone, an approach corridor, and a plume impingement constraint. A contraction metric and feedback gain, computed once offline, provide an explicit exponential convergence rate and a guaranteed tracking tube for every nominal trajectory, without online optimization. Conformal prediction is then used to calibrate the tube radius from Monte Carlo data, which is expected to yield a substantially less conservative radius with
a distribution-free probabilistic guarantee. Future work will extend the
framework to the Tschauner--Hempel equations for docking with targets in
elliptic orbits, where the relative dynamics are time-varying and a
time-varying contraction metric is required.

\section{Acknowledgment}
This work was done as part of the AE 598 Spacecraft Proximity Operations course at University of Illinois, Urbana-Champaign. 
The research ideas, problem formulation, methodology, and expected results presented in this paper are the authors' own. Claude (Anthropic) and OpenAI ChatGPT were used to assist with language editing and with drafting and refining the technical content. All AI-assisted text and mathematical content were reviewed, verified, and revised by the authors.

\bibliographystyle{AAS_publication}   
\bibliography{references}   

\end{document}